\documentclass[reqno]{amsart}
\usepackage{hyperref}
\usepackage{makecell}
\usepackage{multirow}
\usepackage[bb=libus]{mathalpha}
\usepackage{amsmath,amssymb} 
\usepackage[pdftex]{pict2e}
\usepackage{graphicx}
\DeclareGraphicsExtensions{.pdf,.png,.jpg}
\usepackage{bm}
\usepackage[T2A]{fontenc}
\usepackage[utf8]{inputenc}
\usepackage{textcomp}
\usepackage{caption}
\allowdisplaybreaks

\theoremstyle{plain}
\newtheorem{theorem}{Theorem}[section]

\theoremstyle{definition}
\newtheorem{definition}[theorem]{Definition}

\theoremstyle{remark}

\numberwithin{equation}{section}

\usepackage{tikz}
\usepackage{float}

\pdfoutput = 1

\usepackage{euscript}

\usepackage{mathtools}

\usepackage{amsthm}

\begin{document}
	%%%%%%%%%%%%%%%%%%%%%%%%%%%%%%%%%%
		\title[ Exact solution of the generalized Potts model]
	{Exact solution of the three-state generalized double-chain Potts model}

	\author[P. Khrapov]{Pavel Khrapov}
	\address{Pavel Khrapov \\ Department of Mathematics
		\\ Bauman Moscow State Technical University \\  ul. Baumanskaya 2-ya, 5/1, Moscow \\ 105005, Moscow,  Russian Federation}  
	\email{khrapov@bmstu.ru, pvkhrapov@gmail.com }
	
	\author[G. Skvortsov]{Grigory Skvortsov}
	\address{Grigory Skvortsov\\ Department of Mathematics
		\\ Bauman Moscow State Technical University \\  ul. Baumanskaya 2-ya, 5/1, Moscow \\ 105005, Moscow,  Russian Federation}  
	\email{skvortsovga@student.bmstu.ru, skvortsov.grisha@list.ru}
	
	

		\begin{abstract}   
		An exact analytical solution of generalized three-state double-chain Potts model with multi-spin interactions which are invariant under cyclic shift of all spin values is obtained. The partition function in a finite cyclically closed strip of length L, as well as the free energy, internal energy, entropy and heat capacity in thermodynamic limit are calculated using transfer-matrix method. Partial magnetization and susceptibility are suggested as the generalization of usual physical characteristics of a system.\\ \indent Proposed model can be interpreted as a generalized version of standard Potts model (which has Hamiltonian expressed through Kronecker symbols) and clock model (with Hamiltonian expressed through cosines). 
		Considering a particular example of the model with plenty of forces, model's ground states are found, figures of its thermodynamic characteristics and discussed their behaviour at low temperature are shown.
		\end{abstract}
			\maketitle
	
	\tableofcontents
\newpage

	\section{Introduction} 
	\indent Lattice spin models, which originally appeared to explain ferromagnetic phenomena, have now been successfully used to study collective behavior in general \cite{Niss}. For instance, in \cite{Silva} 21-state Potts model was used to study the protein folding process. Generalized Potts models with large number of states appear in many areas, including physics (site percolation in the lattice gas \cite{pottslattice}), biology (polymer gelation \cite{pottspolymer}, studying a cancerous tumor \cite{cancer}), algebra, computer science, sociology, medicine and so on (see \cite{Rozikov},\cite{review}).\\
\indent An important role in the development of the phase transitions theory was played by exact analytical solutions obtained for a few number of lattice spin models. Important technical improvement in the mathematical apparatus of lattice spin models was made in 1941 \cite{Kramers}, when the Ising model was formulated on matrix language and therefore calculation of a partition function was reduced to the search of transfer-matrix's eigenvalues.\\
\indent The Ising model is the Potts model with two states. In 1944 L. Onsager \cite{FullOnsager} obtained an analytical solution of the Ising model with nearest-neighbours (i.e. with two coupling constants: horizontal $J_h$ and vertical $J_ v$ ones) on two-dimensional square lattice. Phase transition temperature of this model was first discovered by Kramers and Wannier \cite{Kramers}:  $$(e^K-1)(e^L-1)=2,$$ where $K=\frac{J_h}{k_B T}, L=\frac{J_v}{k_B T},$ and Potts, generalizing the formula above, showed, that the Potts model has a phase transition at $(e^K-1)(e^L-1)=q,$ where $q$ is the number of states \cite{Potts}. \\
\indent There is a variety of Potts models: standard, generalized, clock, Ashkin-Teller, etc. Exact solutions were obtained for 3-state and 4-state single-chain Potts models with nearest-neighbours \cite{Kassan-Ogly}, for special cases of 6-state  \cite{Kassan-Ogly4} and 12-state \cite{Kassan-Ogly2} models. In \cite{Yurischev} an exact analytical solution for double-chain Potts model with 10 forces in a unit cell and arbitrary integer $q$ was obtained. In \cite{Khrapov} disorder solutions for generalized Potts model were calculated.\\
\indent Potts model can be investigated using graph theory methods (see \cite{Tutte2},\cite{Sokal}), usually with the help of Tutte polynomials. In 1972 Fortuin and Kasteleyn \cite{Kasteleyn} showed the relationship between Potts model and Tutte polynomials and introduced the random cluster model which generalizes Potts model on arbitrary positive non-integer $q$. In \cite{Shrock},\cite{Shrock3} Potts models with different boundary conditions were considered, in \cite{Shrock2} some exact results on Potts model with external field were obtained. In \cite{Ananikyan} authors derived an exact solution of gauge $\mathcal{Z}_4$ Potts model on square and triangular lattices. Cluster properties and bound states of the Yang-Mills model with compact Abelian gauge group were studied in \cite{Khrapov4}.\\
\indent Recently, papers have appeared in which the Ising and Potts models are investigated experimentally:
using non-equilibrium quantum condensators \cite{Berloff}, simulator of quantum computer \cite{Andreev}.
Potts model can be investigated on one-, two-, and three-dimensional lattices: square, triangular, kagome, body-centered cubic, Bethe and honeycomb ones \cite{Shrock5}. There are works in which authors calculate zeros of partition function  \cite{Ananikyan2}, \cite{Sokal2}, correlation functions \cite{Filippov}, investigate ground states \cite{Kashapov} and so on. \\
\indent Potts model is used in many fields, moreover, researchers often need non-standard Potts models with quite specific Hamiltonians (see \cite{Rozikov}, p. 53).
That is why obtaining an exact solution for wide class of Hamiltonians is an important problem. \\
\indent In this paper, three-state double-chain Potts model with cyclic boundary conditions is considered. An exact analytical solution of this model in a finite strip of length $L$ is obtained using transfer-matrix method and analytical expressions for its physical characteristics in thermodynamic limit are derived using Cardano formula (see section \ref{appendix} and \cite{Tabachnikov}).\\
\indent In section \ref{2} the model is introduced, its Hamiltonian is formulated through Kronecker symbols, and main results are listed. Some generalization of Potts model's characteristics such as the partial magnetization and susceptibility are presented. These generalized characteristics allow to represent the mean value of an arbitrary random function of the spin variable $\langle \phi(\sigma) \rangle$, where $\sigma$ is some spin of the lattice, as well as the covariances of such functions in the form of their linear combinations.
\\
\indent In section \ref{3} the clock version of the original Hamiltonian is introduced.\\
\indent In section \ref{4} an example of model given being analyzed in detail: plots of free energy, entropy, heat capacity, internal energy and partial susceptibility are shown, a table of ground states is provided.\\
\indent Section \ref{5} contains proof of theorem \ref{teor}.\\
\indent In Appendix the Cardano formula for solving an equation of the third degree is discussed.

\newpage
	\section{Generalized Potts model with multi-spin interactions invariant under cyclic shift of all spin values} \label{2}
	\begin{figure} [h!] \label{graph}
	\centering
	\includegraphics [width=0.75\textwidth] {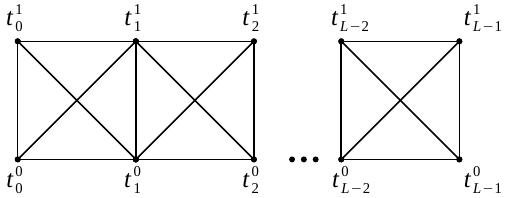}
	\caption{Lattice $\mathcal{L}_{2,L}$, only two-site couplings of the model are displayed. Transfer-matrix $\Theta$ propagates from left to right.}
	\end{figure} 
Consider cyclically closed double-chain Potts model in a finite strip of legth $L:$ on the lattice  $\mathcal{L}_{2,L}:$ $$\ \mathcal{L}_{2,L}=\{t_i^j \text{ } \vert \  i = 0,1, \dots , L-1, \ t_L^j \equiv t_0^j,  j = 0,1\}.$$ Each point $t_i^j$ contains a particle. The state of the particle is determined by the value of the spin $\sigma_i^j \in X=\{ 0, 1, \dots, q-1\},\  \sigma \in X^{\mathcal{L}_{2,L}}$ and the Hamiltonian of $\mathcal{L}_{2,L}$ has the form:
\begin{equation} \label {Hamiltonian_H_L}
	\mathcal{H}_{L} \left(\sigma\right)=\sum_{i=0}^{L-1} H \left( \sigma_i^0, \sigma_i^1; \sigma_{i+1}^0, \sigma_{i+1}^1 \right), 
\end{equation}
Elementary Hamiltonian $H$ of a cell $\Omega_i = \{ t_i^0, t_i^1, t_{i+1}^0, t_{i+1}^1 \}$ takes the form:
\begin{multline} \label{Hamiltonian_H_i}
 H \left(\sigma \vert_{\Omega_i}\right)=H \left( \sigma_i^0, \sigma_i^1; \sigma_{i+1}^0, \sigma_{i+1}^1 \right)=\\
 -\left(\sum_{\{\mu_1,\mu_2,\mu_3\} \in X^3} J_{\mu_1,\mu_2,\mu_3} \delta_{\sigma_0,\mu_1+\sigma_1,\mu_2+\sigma_2,\mu_3+\sigma_3}+\sum_{\mu \in X} h_{\mu} \sum_{i=0}^3\delta_{\mu,\sigma_i} \right)
	\end{multline}
where $\sigma_0 \equiv \sigma_i^0, \sigma_1 \equiv \sigma_i^1, \sigma_2 \equiv \sigma_{i+1}^0, \sigma_3 \equiv \sigma_{i+1}^1,$
  \begin{eqnarray}  \label {chi}
\delta_{ a_1, a_2, \dots, a_s}  = \left\{
                                \begin{array}{ll}
                                  1, & \hbox{if $a_r \equiv a_l (\text{mod } q), r=1,2, \dots s, l=1, \dots s$ } \\
                                  0, & \hbox{otherwise.}
                                \end{array}
                              \right.
\end{eqnarray}
$\delta_{ a_1, a_2, \dots, a_s}$  denotes the Kronecker symbol and the last term of Hamiltonian (\ref{Hamiltonian_H_i}) corresponds to the external field $\vec{h}=\{h_0,h_1,h_2\}$. \\
\indent Note that in further transformations, the Boltzmann constant $k_B$ is set equal to unity, while temperature $T$ and interactions' constants $J_{\mu_1,\mu_2,\mu_3}$ will be measured in
the same units as is usually done in the theory of low-dimensional systems.\newpage
\indent Let us introduce the transfer-matrix $\Theta$ with matrix elements \label{transfer}\\ 
	 $\langle \sigma_0,\sigma_1 \vert \Theta \vert \sigma_2,\sigma_3 \rangle = exp \left[ H(\sigma_0,\sigma_1;\sigma_2,\sigma_3)/T \right]$ and such a structure (see Figure \hyperref[structure]{2}):
	 	 \begin{figure}[h] \label{structure}
	 		\begin{picture}(156,165) 
		\put(156,10){\line(-1,0){156}}
		\put(0,10){\line(0,1){144}}
		\put(12,10){\line(0,1){120}}
		\put(24,10){\line(0,1){144}}
		\put(36,22){\line(0,1){120}}
		\put(48,22){\line(0,1){120}}
		\put(60,22){\line(0,1){132}}
		\put(72,22){\line(0,1){120}}
		\put(84,22){\line(0,1){120}}
		\put(96,22){\line(0,1){132}}
		\put(108,22){\line(0,1){120}}
		\put(120,22){\line(0,1){120}}
		\put(132,22){\line(0,1){132}}
		\put(156,10){\line(0,1){144}}
		\put(156,22){\line(-1,0){156}}
		\put(156,34){\line(-1,0){144}}
		\put(156,46){\line(-1,0){144}}
		\put(156,58){\line(-1,0){156}}
		\put(156,70){\line(-1,0){144}}
		\put(156,82){\line(-1,0){144}}
		\put(156,94){\line(-1,0){156}}
		\put(156,106){\line(-1,0){144}}
		\put(156,118){\line(-1,0){144}}
		\put(156,130){\line(-1,0){156}}
		\put(156,142){\line(-1,0){132}}
		\put(14,14){\tiny{$\sigma_0$}}
		\put(2,14){\tiny{$\sigma_1$}}
		\put(14,26){\small{$2$}}
		\put(14,38){\small{$1$}}
		\put(14,50){\small{$0$}}
		\put(14,62){\small{$2$}}
		\put(14,74){\small{$1$}}
		\put(14,86){\small{$0$}}
		\put(14,98){\small{$2$}}
		\put(14,110){\small{$1$}}
		\put(14,122){\small{$0$}}
		\put(2,37){\small{$2$}}
		\put(2,73){\small{$1$}}
		\put(2,109){\small{$0$}}
		\put(134,134){\tiny{$\sigma_2$}}
		\put(134,146){\tiny{$\sigma_3$}}
		\put(26,134){\small{$0$}}
		\put(38,134){\small{$1$}}
		\put(50,134){\small{$2$}}
		\put(62,134){\small{$0$}}
		\put(74,134){\small{$1$}}		
		\put(86,134){\small{$2$}}
		\put(98,134){\small{$0$}}
		\put(110,134){\small{$1$}}
		\put(122,134){\small{$2$}}
		\put(0,154){\line(1,0){156}}
		\put(38,145){\small{$0$}}
		\put(74,145){\small{$1$}}
		\put(110,145){\small{$2$}}
		\end{picture}
		\caption{The structure of the transfer-matrix $\Theta$}
	 \end{figure} \\
	Then the partition function $$ Z_{L} = \sum_{\sigma \in X^{G_L}} e^{-\frac{\mathcal{H}_L \left( \sigma \right)}{T} }$$ can be represented as $$Z_{L} = \text{Tr}\left( \Theta^L \right).$$
	Let $q=3$ and the external field is equal to zero, i.e. spins $\sigma^i_j$ take three values $0,1,2$ and $\vec{h}=\vec{0}.$ Then Hamiltonian (\ref{Hamiltonian_H_i}) is invariant under cyclic shift of all spin values:
\begin{equation} \label{sd}
H(i+1,j+1; k+1, l+1)=H(i, j; k, l),
\end{equation}   wherein addition is performed in $\mathbb{Z}_3$.\\
 Corresponding transfer-matrix takes the form:
\begin{equation}\label{vid3}
\Theta = \begin {psmallmatrix} 
a& d& e& f& g& h& i& j& k& \\
l& b& m& n& o& p& q& r& s& \\
t& u& c& v& w& x& y& z& A& \\
A& y& z& c& t& u& x& v& w& \\
k& i& j& e& a& d& h& f& g& \\       
s& q& r& m& l& b& p& n& o& \\
o& p& n& r& s& q& b& m& l& \\
w& x& v& z& A& y& u& c& t& \\
g& h& f& j& k& i& d& e& a& \\
\end{psmallmatrix} 	
\end{equation}	 
Due to condition (\ref{sd}) transfer-matrix $\Theta$ (\ref{vid3}) commutes with the permutation matrix $\varkappa:$ 
\begin{equation}\label{kap}
 \varkappa =\begin {psmallmatrix}
0& 0& 0& 0& 1& 0& 0& 0& 0& \\
0& 0& 0& 0& 0& 1& 0& 0& 0& \\
0& 0& 0& 1& 0& 0& 0& 0& 0& \\
0& 0& 0& 0& 0& 0& 0& 1& 0& \\
0& 0& 0& 0& 0& 0& 0& 0&1& \\
0& 0& 0& 0& 0& 0& 1& 0& 0& \\
0& 1& 0& 0& 0& 0& 0& 0& 0& \\
0& 0& 1& 0& 0& 0& 0& 0& 0& \\
1& 0& 0& 0& 0& 0& 0& 0& 0& \\
\end{psmallmatrix} 	
\end{equation}	
This circumstance allows us to find all its eigenvalues. \\
\begin{theorem}\label{teor}
\textit{Partition function of considered model in a cyclically closed finite strip of length $L$ can be presented as :
$$ Z_{L} = \sum_{\varepsilon \in E}\sum_{k = 1}^{3}\lambda_{\varepsilon,k}^L,$$
 where $E = \{ 1,e^{\frac{2 \pi \mathbb{i}}{3}},e^{\frac{4 \pi \mathbb{i}}{3}}\}$ is the set of cubic roots of 1, $\lambda_{\varepsilon,k}$ are eigenvalues of matrices:}\\
\begin{equation} \label{sdd}
\Theta_{\varepsilon}=\left(
\begin{array}{ccc}
 a+g \varepsilon +k \varepsilon ^2 & d+h \varepsilon +i \varepsilon ^2 & e+f \varepsilon +j \varepsilon ^2 \\
 l+o \varepsilon +s \varepsilon ^2 & b+p \varepsilon +q \varepsilon ^2 & m+n \varepsilon +r \varepsilon ^2 \\
 t+w \varepsilon +A \varepsilon ^2 & u+x \varepsilon +y \varepsilon ^2 & c+v \varepsilon +z \varepsilon ^2 \\
\end{array}
\right)
\end{equation}
The principal (single largest real) eigenvalue of the matrix $\Theta_1$ determined from its secular equation, can be expressed in radicals (see \ref{appendix}) and coincides with the prinicpal eigenvalue of transer-matrix $\Theta.$
Furthermore, $\overline{\lambda}_{\varepsilon,k}=\lambda_{\overline{\varepsilon},k}, k=1,2,3.$
\end{theorem}
  \indent Main models's thermodynamic characteristics in a strip of length $L$ can be expressed through its partition function  \cite{Baxter}:
	\begin{itemize}
	\item Free energy is equal to:
	\begin{equation} \label{nn1}
		f(T) = - \frac{T}{N} \ln Z_{L}(T),
	\end{equation}
	where $N=2L$ - the number of lattice's sites.\\
	\item Internal energy is equal to:
	\begin{equation} \label{nnn2}
		u(T) = -T^2 \frac{\partial}{ \partial T}\biggl(\frac{f(T)}{T}\biggr). 
	\end{equation}
	\item Entropy is expressed as:
	\begin{equation} \label{last9}
		S(T)=-\frac{\partial}{\partial T} f(T),
	\end{equation}
	\item Heat capacity is:
	\begin{equation} \label{nnn3}
		c(T) = \frac{\partial}{\partial T}u(T) = -2T\frac{\partial}{\partial T}\biggl(\frac{f(T)}{T}\biggr) - T^2 \frac{\partial^2}{\partial T^2}\biggl(\frac{f(T)}{T}\biggr).
	\end{equation}
	\end{itemize}
\begin{definition}	    \indent \ Let random variable $\mathcal{M_{\mu}}$ be $\mathcal{M_{\mu}}=\sum\limits_{i=0}^{L-1}\sum\limits_{j=0}^1 \delta_{\mu,\sigma_i^j}.$ \\
\end{definition}
\begin{definition}
	 Partial magnetization $m_{\mu}$ and partial susceptibility $\chi_{\mu_0,\mu_1}$ of double-chain Potts model are determined as:
	\begin{equation}
	 m_{\mu}=\lim_{L \rightarrow \infty} \frac{1}{2L} \langle \mathcal{M_{\mu}} \rangle
	 \end{equation}
	\begin{equation}\label{parc}
	 \chi_{\mu_0,\mu_1}= \lim_{L \rightarrow \infty} \frac{\langle \mathcal{M}_{\mu_0} \mathcal{M}_{\mu_1} \rangle - \langle \mathcal{M}_{\mu_0} \rangle \langle \mathcal{M}_{\mu_1}\rangle}{2LT} =\lim_{L \rightarrow \infty} \frac{cov(\mathcal{M}_{\mu_0},\mathcal{M}_{\mu_1})}{2LT} 
	 \end{equation}
    These quantities can be calculated by derivation of free energy:	
	\begin{equation} \label{last1}
	 m_{\mu} (\vec{h},T) =-\frac{\partial}{\partial h_{\mu}} f(\vec{h},T) \
	\end{equation}
	\begin{equation} \label{last2}
	 \chi_{\mu_0,\mu_1} = -\frac{\partial^2}{\partial h_{\mu_0} \partial h_{\mu_1}} f(\vec{h},T)
	\end{equation}
\end{definition}
	\begin{theorem} \label{th1}
		\textbf{\textit{Main theorem}}\\
		\indent\textit{For generalized double-chain three-state Potts models with Hamiltonian (\ref{Hamiltonian_H_L}) - (\ref{Hamiltonian_H_i}) and zero external field $\vec{h}=\vec{0}$ in thermodynamical limit (i.e. at $ L \rightarrow \infty$) free energy, internal energy, entropy, heat capacity, partial magnetizations and susceptibilities can be expressed through $\lambda_{max}(T)$, i.e. the principal eigenvalue of matrix $\Theta_1$:}
		\begin{equation} \label{16}
			f(T) = -T\frac{\ln(\lambda_{max}(T))}{2},
		\end{equation}
		\begin{equation} \label{17}
			u(T) =  T^2 \frac{\partial}{\partial T} \frac{\ln(\lambda_{max}(T))}{2},
		\end{equation}
		\begin{equation} \label{last10}
			S(T) = \frac{1}{2} \biggl(\ln (\lambda_{max}(T)) + \frac{T}{\lambda_{max}(T)} \frac{\partial \lambda_{max}(T)}{\partial T}\biggr),
		\end{equation}
		\begin{equation} \label{18}
			c(T) = 2T \frac{\partial}{\partial T} \frac{\ln (\lambda_{max}(T))}{2} + T^2 \frac{\partial^2}{\partial T^2}\frac{\ln (\lambda_{max}(T))}{2},
		\end{equation}
		\begin{equation} \label{last3}
			m_{\mu}(\vec{h},T) \bigg\rvert_{\vec{h}=\vec{0}} = \frac{T}{2} \frac{1}{\lambda_{max}(T)} \frac{\partial \lambda_{max}(\vec{h},T)}{\partial h_{\mu}} \bigg\rvert_{\vec{h}=\vec{0}} =\frac{1}{3},            
		\end{equation}
{\small		\begin{equation} \label{last4}
\begin{gathered}
			\chi_{\mu_0, \mu_1}(\vec{h},T) \bigg\rvert_{\vec{h}=\vec{0}} =\biggl\{  \frac{T}{2}\biggl(\frac{1}{\lambda_{max}(T)}\frac{\partial^2 \lambda_{max}(\vec{h},T)}{\partial h_{\mu_0} \partial h_{\mu_1}} - \\
			-\frac{1}{\lambda^2_{max}(T)} \frac{\partial \lambda_{max}(\vec{h},T)}{\partial h_{\mu_0}} \frac{\partial \lambda_{max}(\vec{h},T)}{\partial h_{\mu_1}} \biggr) \biggr\} \bigg\rvert_{\vec{h}=\vec{0}}
			\end{gathered}
		\end{equation}}
		
		\textit{where partial derivatives of $\lambda_{max}(\vec{h}, T)$ has the form:}
			\begin{equation} \label{last11}			
\frac{\partial \lambda_{max}(\vec{h},T)}{\partial h_{\mu}} = -\frac{\sum\limits_{n=0}^9 \frac{\partial a_n}{\partial h_{\mu}}  \lambda_{max}^n}{\sum\limits_{n=0}^8(n+1)a_{n+1}\lambda_{max}^n},					
		\end{equation}
{\small			\begin{multline} \label{last6}
			\frac{\partial^2 \lambda_{max}(\vec{h},T)}{\partial h_{\mu_0} \partial h_{\mu_1}} =\frac{-1}{\sum \limits_{n=0}^8 (n+1)a_{n+1} \lambda_{max}^n} \biggl(
\sum\limits_{n=0}^9 \frac{\partial^2 a_n}{\partial h_{\mu_0} \partial h_{\mu_1}}  \lambda_{max}^n + \\ 
 \frac{\partial \lambda_{max}}{\partial h_{\mu_1}}  \sum\limits_{n=0}^8 (n+1) \frac{\partial a_{n+1}}{\partial h_{\mu_0}}  \lambda_{max}^n + \\
+\frac{\partial \lambda_{max}}{\partial h_{\mu_0}} \left[ \sum\limits_{n=0}^8 (n+1) \frac{\partial a_{n+1}}{\partial h_{\mu_1}}  \lambda_{max}^n + \frac{\partial \lambda_{max}}{\partial h_{\mu_1}} \biggl( \sum\limits_{n=0}^7 (n+2)(n+1) a_{n+1} \lambda_{max}^n   \biggr) \right]\biggr)			
		\end{multline}}
	\end{theorem}
	\subsection{Сommentary}   In this paper, thermodynamic characteristics were found only at zero external field. It is clear that in this case (see (\ref{sd})) partial magnetizations $m_{\mu}=\frac{1}{3}$, because $m_{\mu_1}=m_{\mu_2}$ and $\sum\limits_{\mu=0}^2 m_{\mu}=1.$\\
\indent Formulae (\ref{last11}), (\ref{last6}) are obtained by taking the derivative of the transfer-matrix's $\Theta$ secular equation: $ \sum\limits_{n=0}^9 a_n \lambda^n=0$ with respect to $h_{\mu}$. In a similar way, one can differentiate secular equation with respect to $T$ and obtain analogous formulae for entropy, internal energy and heat capacity as is done in \cite{Volkov}.
 \section{Generalized clock Potts model at $q=3$} \label{3}\indent \\
\indent Let us rewrite the elementary Hamiltonian (\ref{Hamiltonian_H_i}) with zero external field  $\vec{h}=\vec{0}$ in the form of generalized clock Potts model's Hamiltonian:
{\small
 	\begin{multline} \label{Hamcos}	
		H(\sigma_0,\sigma_1;\sigma_2,\sigma_3)
		=-\biggl[ A_1 \cos \left(\frac{2\pi}{3}   (\sigma_1-\sigma_0)\right)+ A_2 \cos \left(\frac{2\pi}{3}   (\sigma_2-\sigma_0)\right)+\\
		A_3 \cos \left(\frac{2\pi}{3}   (\sigma_3-\sigma_0)\right)+A_4 \cos \left(\frac{2\pi}{3}  (\sigma_3-\sigma_1)\right)+A_5 \cos \left(\frac{2\pi}{3}   (\sigma_2-\sigma_1)\right)+\\
		A_6 \cos \left(\frac{2\pi}{3}  (\sigma_3-\sigma_2)\right)+\tilde{A}_1 \sin \left(\frac{2\pi}{3} (\sigma_1-\sigma_0)\right)+\tilde{A}_2 \sin \left(\frac{2\pi}{3}  (\sigma_2-\sigma_0)\right)+\\
		\tilde{A}_3 \sin \left(\frac{2\pi}{3} (\sigma_3-\sigma_0)\right)+\tilde{A}_4 \sin \left(\frac{2\pi}{3}   (\sigma_3-\sigma_1)\right)+\tilde{A}_5 \sin \left(\frac{2\pi}{3}  (\sigma_2-\sigma_1)\right)+\\
		\tilde{A}_6 \sin \left(\frac{2\pi}{3} (\sigma_3-\sigma_2)\right)+B_1 \cos \left(\frac{2\pi}{3}   (\sigma_0+\sigma_1+\sigma_2)\right)+B_2 \cos \left(\frac{2\pi}{3}  (\sigma_0+\sigma_1+\sigma_3)\right)+\\
		B_3 \cos \left(\frac{2\pi}{3}  (\sigma_0+\sigma_2+\sigma_3)\right)+B_4 \cos \left(\frac{2\pi}{3}  (\sigma_1+\sigma_2+\sigma_3)\right)+\\
		\tilde{B}_1 \sin \left(\frac{2\pi}{3} (\sigma_0+\sigma_1+\sigma_2)\right)+\tilde{B}_2 \sin \left(\frac{2\pi}{3}  (\sigma_0+\sigma_1+\sigma_3)\right)+\\
		\tilde{B}_3 \sin \left(\frac{2\pi}{3}  (\sigma_0+\sigma_2+\sigma_3)\right)+\tilde{B}_4 \sin \left(\frac{2\pi}{3}  (\sigma_1+\sigma_2+\sigma_3)\right)+\\
		C_1 \cos \left(\frac{2\pi}{3} (\sigma_0-\sigma_1+\sigma_2-\sigma_3)\right)+C_2 \cos \left(\frac{2\pi}{3}  (\sigma_0-\sigma_1-\sigma_2+\sigma_3)\right)+\\
		C_3 \cos \left(\frac{2\pi}{3}  (\sigma_0+\sigma_1-\sigma_2-\sigma_3)\right)+C_4+\tilde{C}_1 \sin \left(\frac{2\pi}{3}   (\sigma_0-\sigma_1+\sigma_2-\sigma_3)\right)+\\
		\tilde{C}_2 \sin \left(\frac{2\pi}{3}  (\sigma_0-\sigma_1-\sigma_2+\sigma_3)\right)+\tilde{C}_3 \sin \left(\frac{2\pi}{3} (\sigma_0+\sigma_1-\sigma_2-\sigma_3)\right) \biggr]
	\end{multline}
	}
	Hamiltonian (\ref{Hamcos}) is equivalent to the original one (\ref{Hamiltonian_H_i}) with $\vec{h}=\vec{0}$,  as both of them cover all functions on $X^4$ (where $X=\{ 0,1,2\}$) which are invariant under cyclic shift of all spin values. That is why results of previous section  (\ref{2}) are valid for such a clock models. The main advantage of clock version of Hamiltonian is that in this form double, triple and quadruple interactions are extracted.\\ \\
	 \begin{minipage}{0.45\textwidth}
 \centering
  \label{123}
\begin{tikzpicture}[node distance={12mm}, thick, main/.style = {draw, circle},main_od/.style = {draw, circle,fill=blue}] 
\node[main] (0) {$\sigma_0$}; 
\node[main] (1) [above of=0] {$\sigma_1$}; 
\node[main] (2) [right of=1] {$\sigma_3$}; 
\node[main] (3) [below of=2] {$\sigma_2$}; 

\draw[very thin] (0) -- (1);\draw[very thin] (2) -- (0);
\draw[very thin] (3) -- (1);\draw[very thin] (0) -- (3);
\draw[very thin] (1) -- (2);
\end{tikzpicture} 

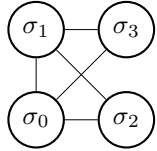
\captionof{figure}{Step of the transfer-matrix $\Theta$, only two-site couplings are displayed}
 \end{minipage}
 \begin{minipage}{0.5\textwidth}
 \indent It should be marked that in further calculations constant $C_4$ is set equal to zero, moreover, transfer-matrix $\Theta$, propagating from left to right, will cover two-site interaction of $\sigma_2$ and $\sigma_3$ in the next step (see Figure \hyperref[123]{3}), that is why we put: $$ A_6=\tilde{A}_6=C_4=0$$
 \end{minipage}   
\newpage
  \section{Examples of physical characteristics} \label{4}
	\label{Example} 
	In order to prevent increasing the number of parameters, constants of double, triple and quadruple forces are usually set equal to each other (see \cite{Shrock},\cite{Wu}). Let us put in Hamiltonian \ref{Hamcos} 
$ A_i \equiv A , \tilde{A}_i \equiv \tilde{A} , i=1, \dots, 5; \ B_j \equiv B , \tilde{B}_j \equiv \tilde{B} ,j=1,  \dots, 4;\   C_k \equiv C , \tilde{C}_k \equiv \tilde{C} ,k=1, 2, 3.$\\
	We will change double interactions:
$A=\tilde{A} = r,\ r \in [-1,1]$ and set $ B=0,C=0.1, \tilde{B}=0.2,\tilde{C}=0.3.$\\
Let us show the free energy graph of this model:
\begin{figure}[H]
				\begin{minipage}[h]{0.9\linewidth}
					\centering{\includegraphics[width=0.7\linewidth]{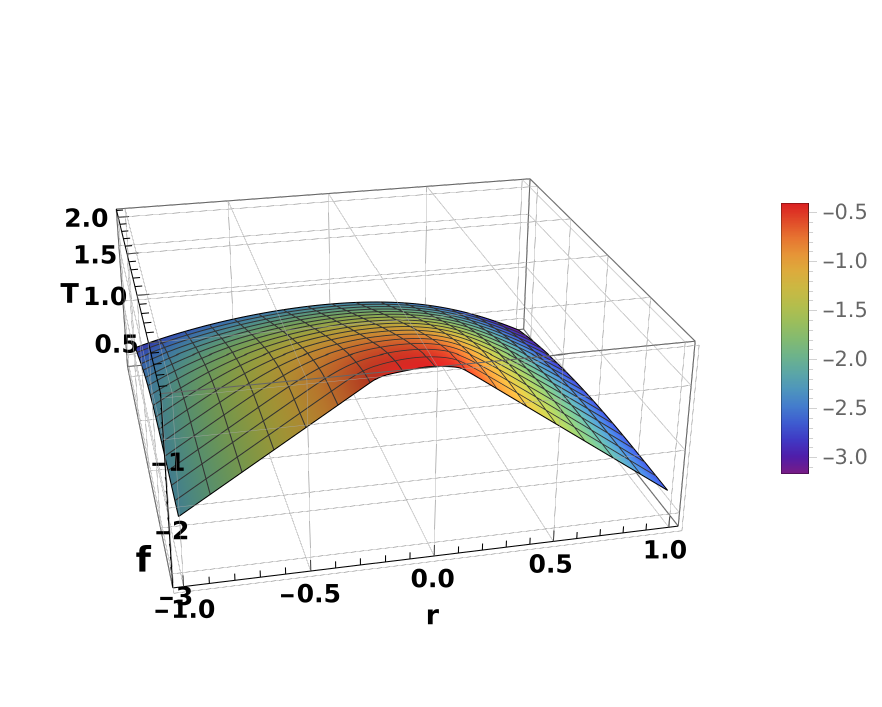}  }
				\end{minipage}
				\begin{minipage}[h]{0.495\linewidth}
					\centering{\includegraphics[width=0.9\linewidth]{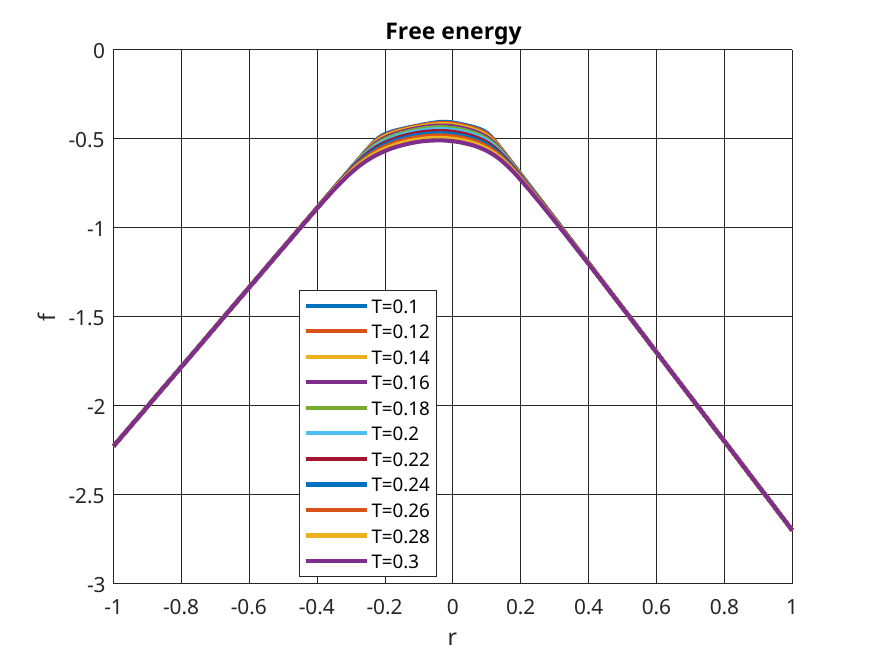}  }
				\end{minipage}
				\begin{minipage}[h]{0.495\linewidth}
					\centering{\includegraphics[width=0.9\linewidth]{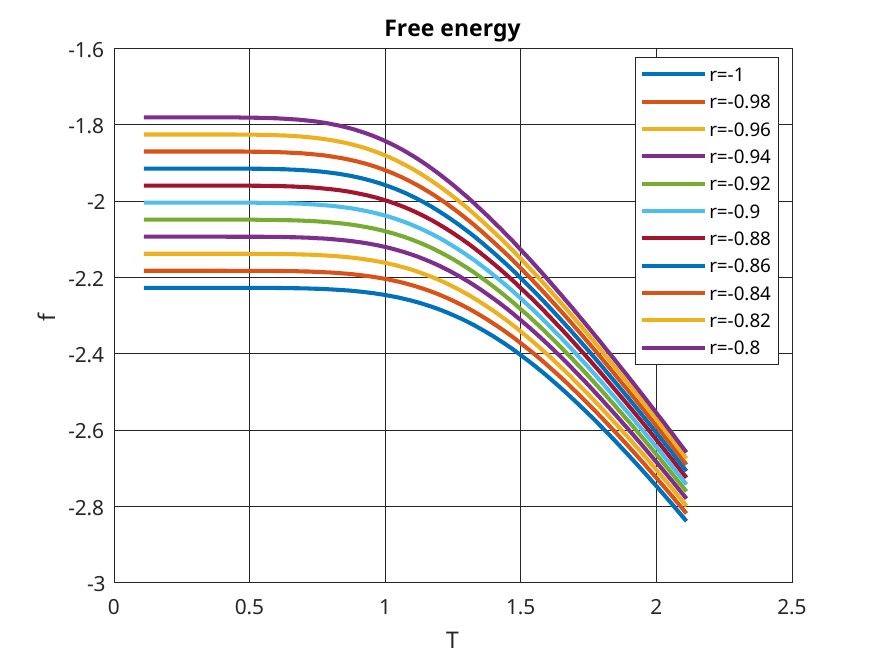} }
				\end{minipage}
				\caption{Plots of free energy, $r \in [-1,1],\  T \in [-0.1;2.1]$}
\end{figure}
The most interesting is model's behaviour at low temperature, i.е. at $T \rightarrow 0+$. It is obvious that $\lim\limits_{T \rightarrow 0+} f(r,T)=\lim\limits_{T \rightarrow 0+} u(r,T).$ In addition (see table \ref{tablica}), one can easily show that at $T \rightarrow 0+$  after replacing the matrix's $\Theta_1$ secular equation coefficients by equivalent infinitely large ones $\lambda_{max}$ will satisfy  $$\lambda^3-e^{-\frac{2 u_1}{T} } \lambda^2 - e^{- \frac{6u_2}{T} }=0 \ \ \ at \ \ \ r \in \left(-\infty;r_2 \right),$$ 
$$\lambda^3- e^{- \frac{6u_2}{T} } - e^{- \frac{6u_3}{T} }=0 \ \ \  at \  \ \ r \in \left(r_1;r_3 \right),$$ 
$$\lambda^3- e^{- \frac{2u_4}{T} } \lambda^2 - e^{- \frac{6u_3}{T} }=0 \ \ \ at \ \ \ r \in \left(r_2;\infty \right),$$
where $u_i, \ i=1, \dots, 4$ denote the limits of internal energy at $T \rightarrow 0+$ (see subsection \ref{equa}).
\subsection{Structure of ground states} \label{equa} \indent \\
			\indent Transfer-matrix \ref{vid3} can be interpreted as a weight matrix of some directed graph with 9 vertices. Obviously, the ground state (i.e. the state with minimal energy per spin) must be periodic, since the number of vertices is finite, аnd therefore the initial vertex must be repeated at some step $L \leq 9$. However, the number of paths of length $L$ starting and ending at the same vertex is finite, hence there exists a path (and its corresponding spin configuration $\sigma$) with minimal energy per spin: $\frac{\mathcal{H}_L \left( \sigma\right)}{2L} \rightarrow min$. Thus, the search of ground states can be reduced to the search of a minimum mean weight cycle in a directed graph. It is clear that such a cycle can be started from any of its elements (compare, for instance, spin configurations at $r \in \left(r_1;r_2 \right)$). Moreover (see (\ref{sd})), states derived from ground ones by cyclic shift of all spin values are also ground states. That is why only one of them is presented (see table \ref{tablica}):
			\renewcommand{\arraystretch}{1.5} 
							\begin{table}[h] 
		\centering
		\caption{Color designation} 	
		\begin{tabular}{| c | c | c | c |} \hline
			\makecell{Color} & \makecell{\centering{\includegraphics[height=0.03\hsize ]{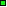}}} & \makecell{\centering{\includegraphics[ height=0.03\hsize]{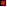}}} & \makecell{\centering{\includegraphics[ height=0.03\hsize]{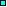}}}  \\ \hline
			\makecell{State} & 0 & 1 & 2   \\ \hline
		\end{tabular}
	\end{table}
			\renewcommand{\arraystretch}{2.5} 
				\begin{table}[h] 
		\centering
		\caption{Ground states} 	\label{tablica}
		\begin{tabular}{| c | c |} \hline
			\makecell{Value of r} & \makecell{Configuration $\sigma$ of ground state}  \\ \hline
			\makecell{$r \in \left(-\infty;r_1=\frac{3 \left(7-13 \sqrt{3}\right)}{10 \left(11 \sqrt{3}+3\right)}\right)$} & \makecell{\centering{\includegraphics[height=0.05\hsize]{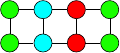}}}   \\ \hline
			\multirow{3}{*}{$r \in \left(r_1;r_2=\frac{3 \left(1-\sqrt{3}\right)}{20 \left(\sqrt{3}+3\right)}\right)$} & \makecell{\centering{\includegraphics[height=0.05\hsize]{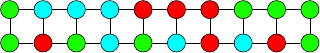}}}  \\  \cline{2-2}
              & \makecell{\centering{\includegraphics[height=0.05\hsize]{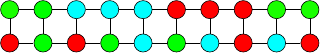}}}   \\		 \cline{2-2}
			  & \makecell{\centering{\includegraphics[height=0.05\hsize]{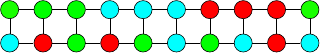}}}   \\ \hline
\multirow{3}{*}{$r \in \left(r_2;r_3=\frac{33 \left(1-\sqrt{3}\right)}{10 \left(\sqrt{3}-27\right)}\right)$} & \makecell{\centering{\includegraphics[height=0.05\hsize]{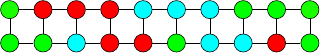}}}  \\   \cline{2-2}
             & \makecell{\centering{\includegraphics[height=0.05\hsize]{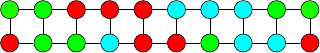}}}   \\  \cline{2-2}		
		 & \makecell{\centering{\includegraphics[height=0.05\hsize]{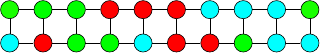}}}   \\ \hline			 
						 
			\makecell{$r \in \left(r_3;\infty\right)$} & \makecell{\centering{\includegraphics[height=0.05\hsize]{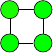}}}   \\ \hline
		\end{tabular}
	\end{table}

\indent \\
Also  ground states at $r_i$ are ignored to avoid making the table \ref{tablica} more cumbersome.\\
\indent As internal energy is the average of Hamiltonian per spin, we obtain that \\ $\lim\limits_{T \rightarrow 0+} u(r,T)=\frac{\mathcal{H}_L \left( \sigma\right)}{2L},$ herewith configuration $\sigma$ can be found in table \ref{tablica} : 
{ \small
\begin{equation*} 
 \lim\limits_{T \rightarrow 0+} u(r,T)=\begin{cases}
u_1= \left(\sqrt{3}+\frac{1}{2}\right) r+\frac{1}{40} \left(3 \sqrt{3}-5\right), \ r \in [-\infty;r_1=\frac{3 \left(7-13 \sqrt{3}\right)}{10 \left(11 \sqrt{3}+3\right)} \approx -0.211]\\
u_2= \frac{1}{60} \left(5 \left(\sqrt{3}+3\right) r-15 \sqrt{3}+3\right), \ r \in [r_1;r_2=\frac{3 \left(1-\sqrt{3}\right)}{20 \left(\sqrt{3}+3\right)} \approx -0.023]     \\
u_3=\frac{1}{120} \left(-10 \left(\sqrt{3}+3\right) r-33 \sqrt{3}+9\right)	, \ r \in [r_2;r_3=\frac{33 \left(1-\sqrt{3}\right)}{10 \left(\sqrt{3}-27\right)} \approx 0.096]\\
u_4=   -\frac{5 r}{2}-\frac{1}{5}, \ r \in [r_3;\infty]
 \end{cases} 
\end{equation*}
One can plot this function and compare it with the plot of internal energy at low temperature (see Figure \hyperref[inlow]{5}):\\
\begin{figure}[h] \label{inlow}
\centering{\includegraphics[width=\linewidth]{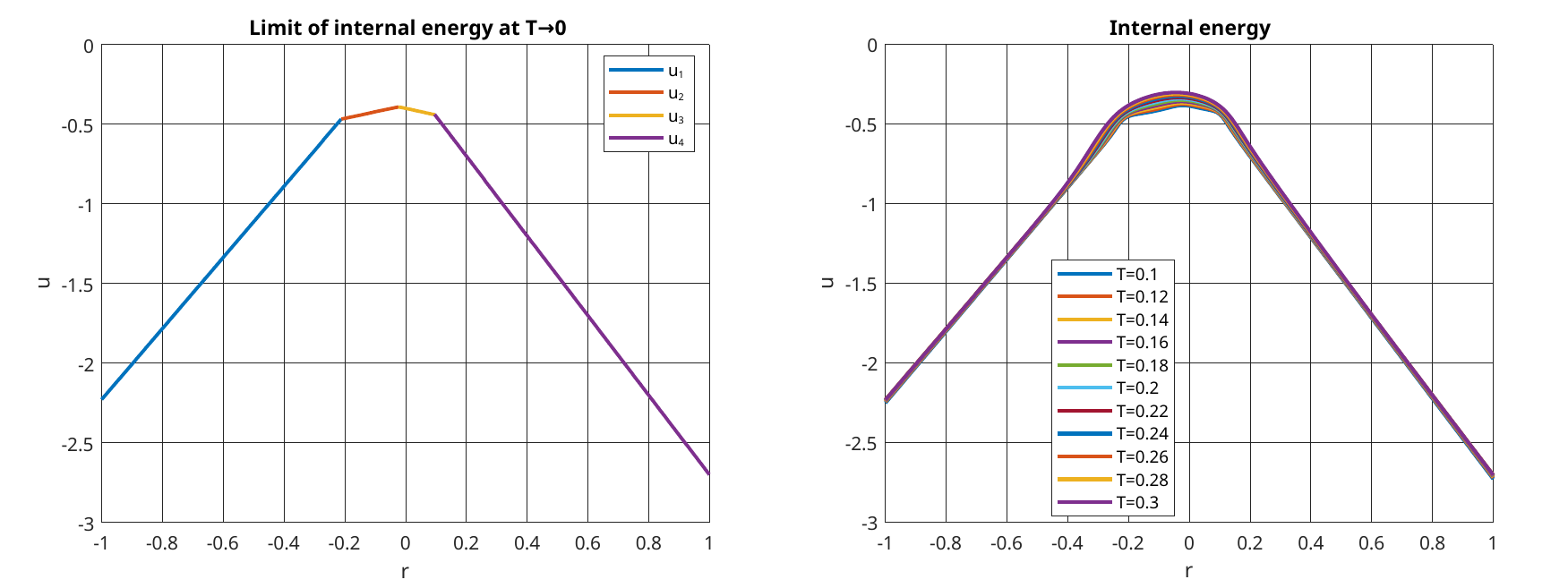}  }
\caption{Plots of internal energy in low-temperature region and its limit at $T \rightarrow 0$, $r \in [-1,1]$}
\end{figure}\\
\begin{figure}[h]
\begin{minipage}[h]{0.495\linewidth}
					\centering{\includegraphics[width=\linewidth]{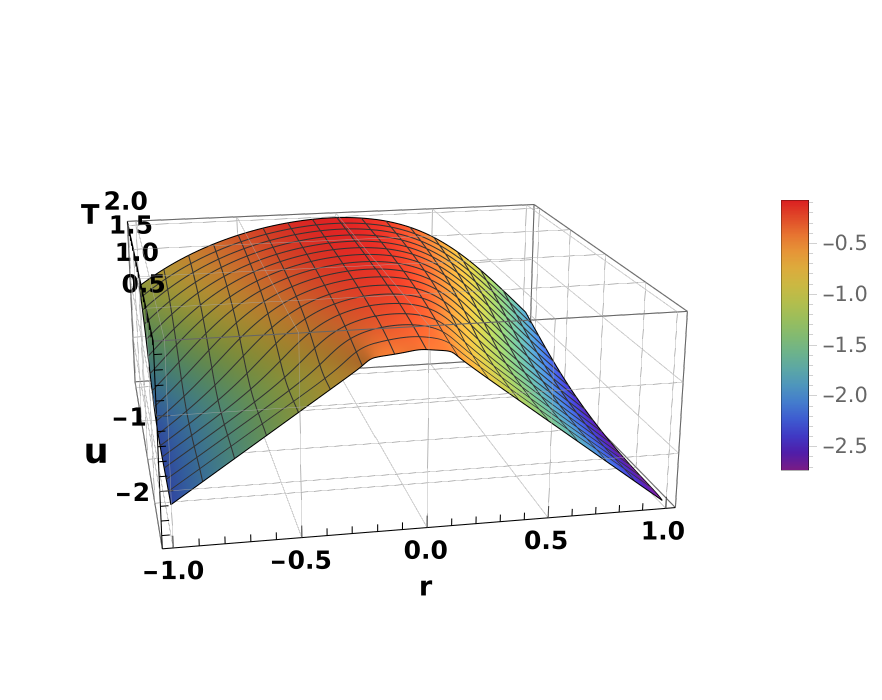} \\  }
				\end{minipage}
				\begin{minipage}[h]{0.495\linewidth}
					\centering{\includegraphics[width=\linewidth]{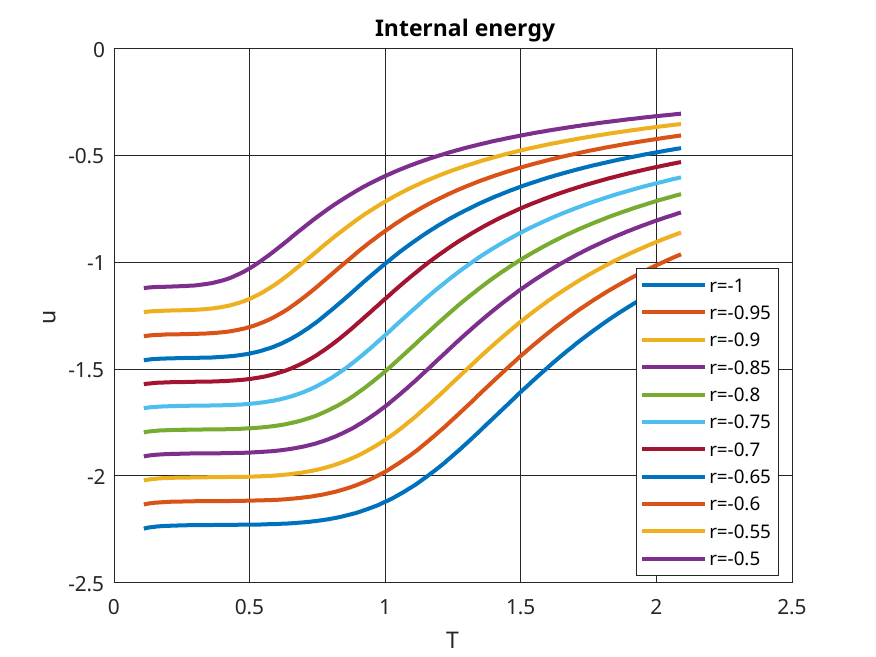} \\ }
				\end{minipage}
				\caption{Plots of internal energy, $r \in [-1,1],\  T \in [-0.1;2.1]$}
\end{figure}

\subsection{Entropy} \indent \\
	\indent At low temperature entropy $S \approx \frac{\ln N'}{2L},$ where $L-$ length of double chain, $N'-$ number of configurations with minimal energy ( i.e. configurations that survive at $T \rightarrow 0+$) and therefore the limit $\lim\limits_{T \rightarrow 0+} S(r,T)$ is equal to zero everywhere, except of three frustration points $r_i$, since the number of ground states at other points is finite (see table \ref{tablica}). At the first and third frustration points, the limit is equal to half of the logarithm of equation's \  $  {\lambda}^3-{\lambda}^2-1=0$ maximum root, that is $$\lim\limits_{T \rightarrow 0+} S(r,T)=\frac{1}{2} \ln \left(\frac{1}{3} \left(\sqrt[3]{\frac{1}{2} \left(3 \sqrt{93}+29\right)}+\sqrt[3]{\frac{1}{2} \left(29-3 \sqrt{93}\right)}+1\right)\right) \approx 0.1911$$
		at $r=r_i,i=1,3.$
		At the second frustration point $\lim\limits_{T \rightarrow 0+} S(r_2,T)=\frac{1}{6} \ln2 \approx 0.1155$\\
		One can see it on the plot (see Figure \hyperref[entlow]{7}):\\
		\begin{figure}[h]\label{entlow}
		\begin{minipage}[h]{0.495\linewidth}
					\centering{\includegraphics[width=\linewidth]{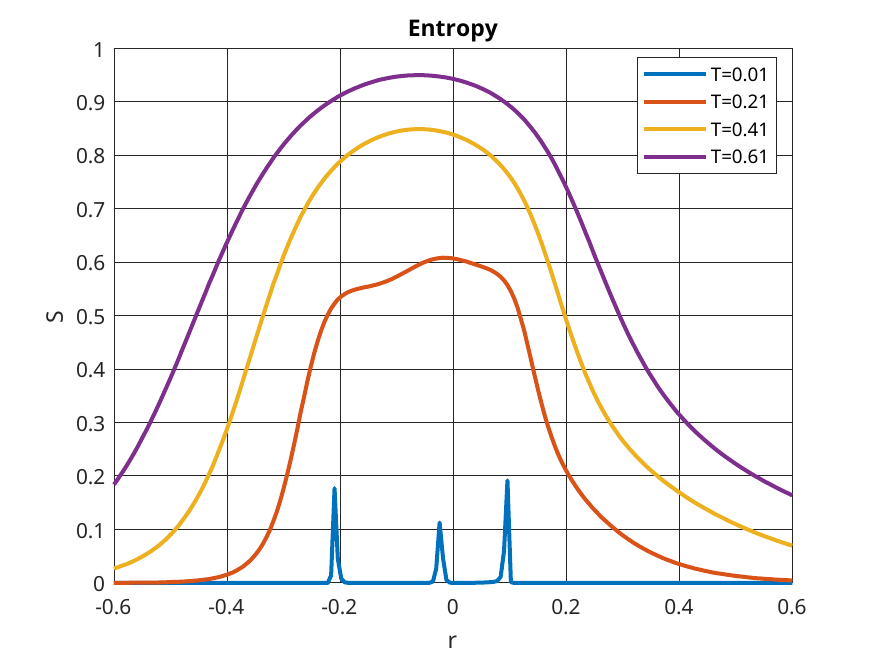} \\ }
				\end{minipage}
				\begin{minipage}[h]{0.495\linewidth}				
				\begin{center}
				\includegraphics[width=\linewidth]{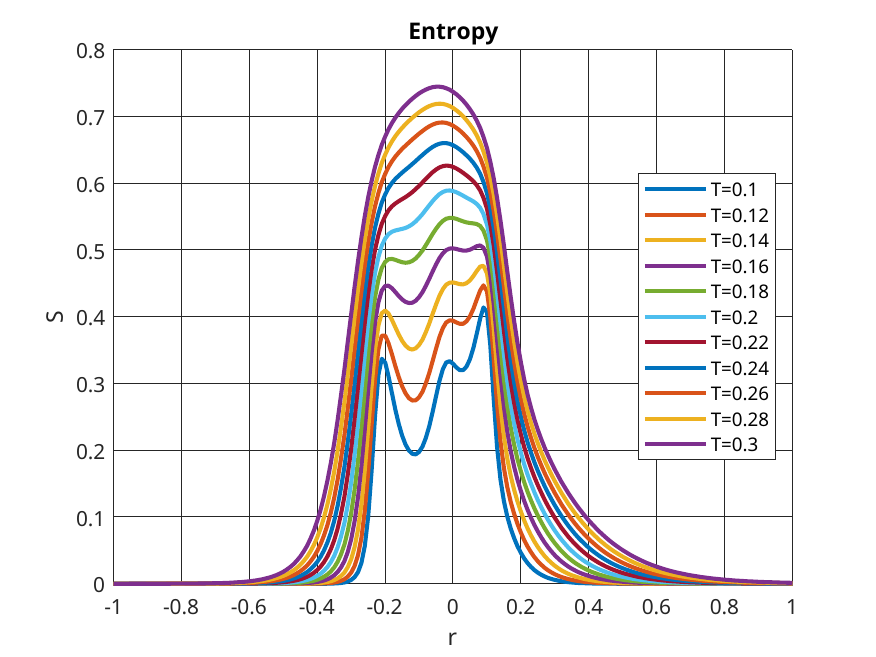}
				\end{center}
				\end{minipage}
				\caption{Plots of entropy in low-temperature region,\  $r \in [-1,1]$}
		\end{figure}
		
				\begin{figure}[h]
					\centering{\includegraphics[width=0.7\linewidth]{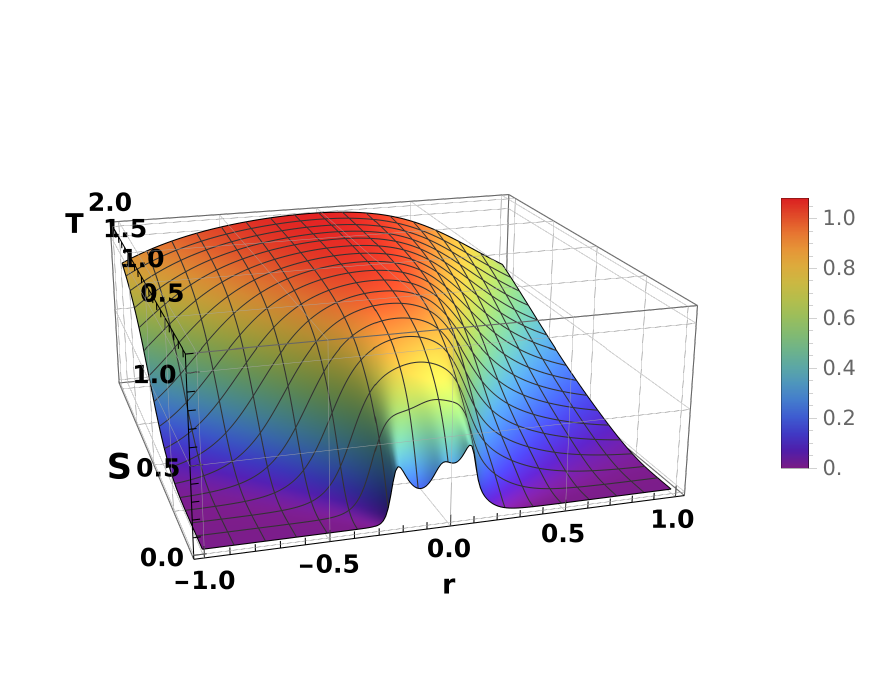}}
					\caption{Plot of entropy, $r \in [-1,1],\  T \in [-0.1;2.1]$ }
				\end{figure}
				
				\begin{minipage}[h]{0.495\linewidth}
				\centering{\includegraphics[width=\linewidth]{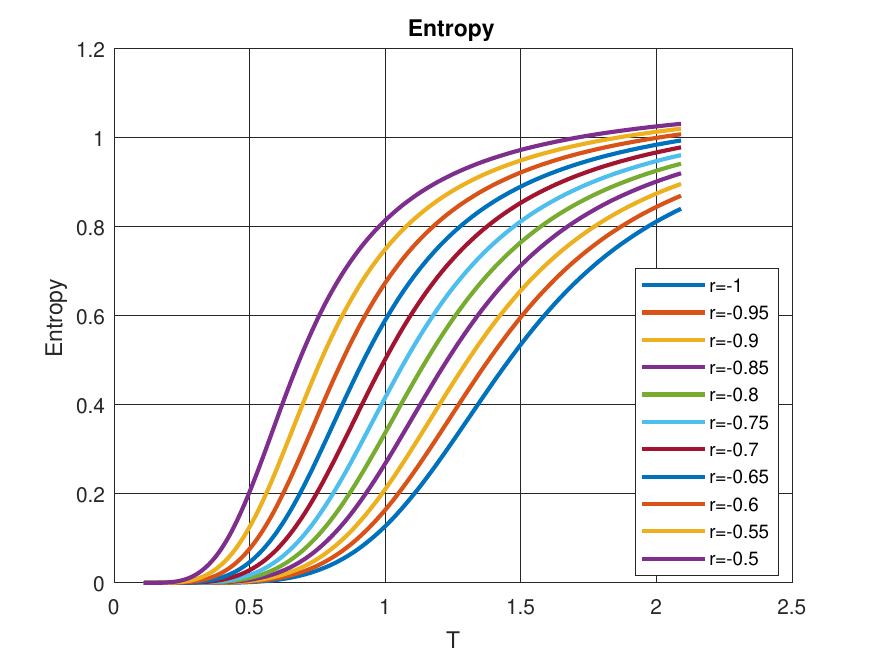}}
				\captionof{figure}{Plot of entropy, $ T \in [-0.1;2.1]$}
				\end{minipage}
					\begin{minipage}[h]{0.495\linewidth}
					Clearly $$\lim\limits_{T \rightarrow \infty} S(r,T)=\ln 3 \approx 1,099$$ because at high temperature all  $9^L$ configurations have similar energy.
				\end{minipage}\\ 
\indent Let us show the plot of heat capacity. In the low-temperature region (see Figure \hyperref[heatcapac]{10}) one can see three sharp minima at frustration points:		
			\begin{figure}[H] \label{heatcapac}
			    \begin{minipage}[h]{0.8\linewidth}
					\centering{\includegraphics[width=\linewidth]{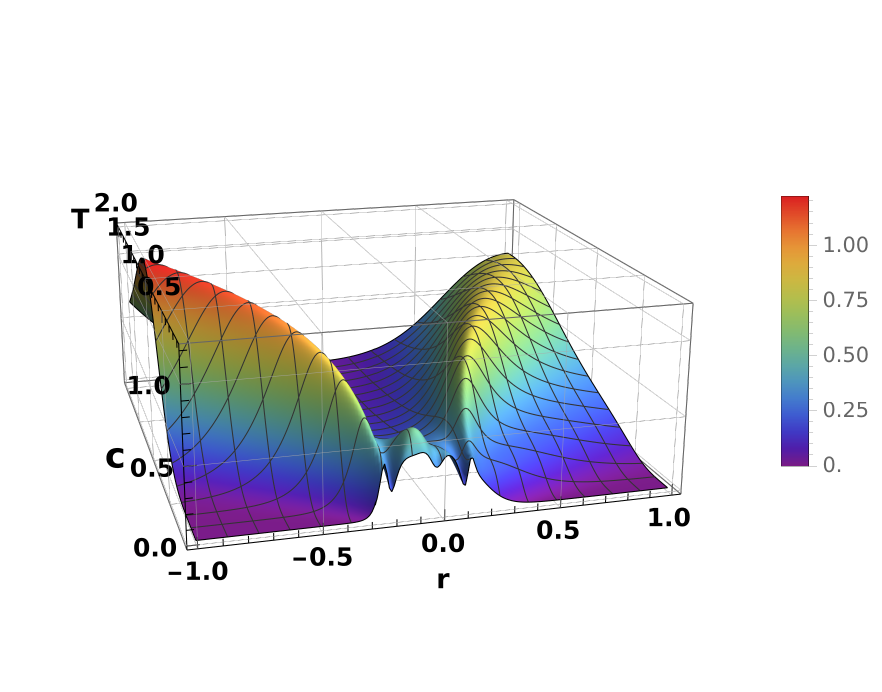} \\  }
				\end{minipage}
				\begin{minipage}[h]{0.495\linewidth}
					\centering{\includegraphics[width=\linewidth]{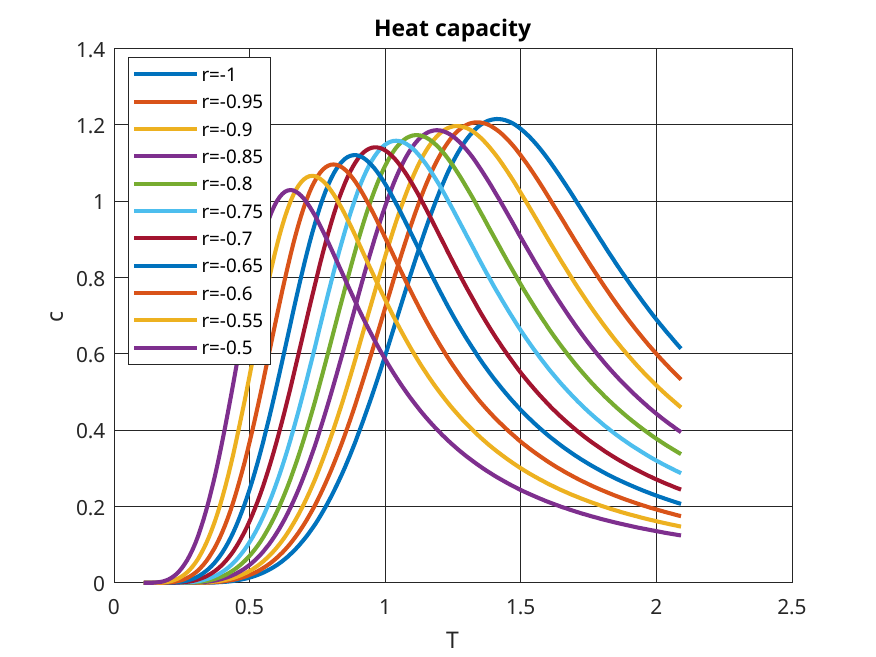} \\ }
				\end{minipage}
				\begin{minipage}[h]{0.495\linewidth}
					\centering{\includegraphics[width=\linewidth]{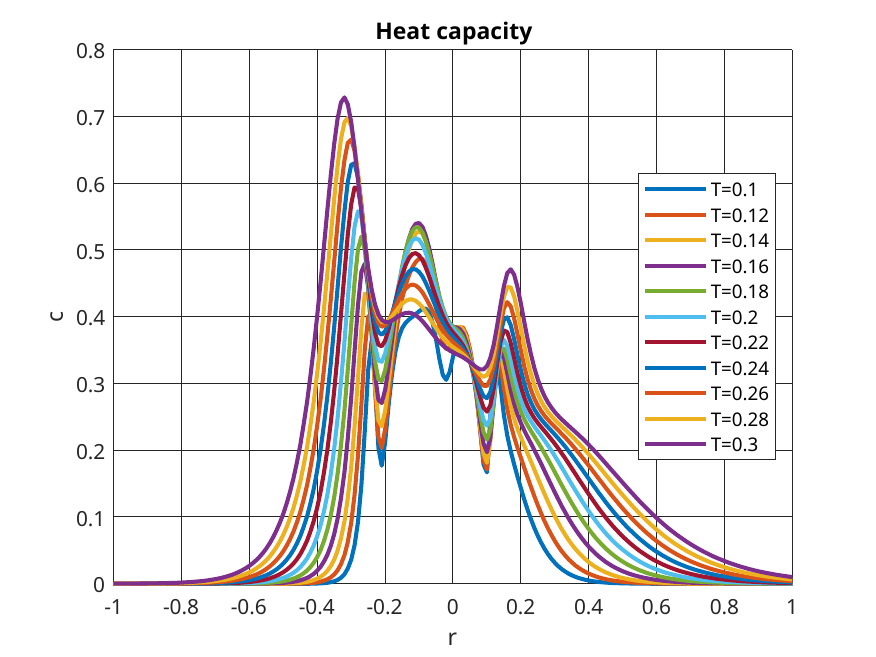} \\ }
				\end{minipage}	
				\caption{Plots of heat capacity,\  $r \in [-1;1],T \in [0.1;2.1]$}			
				\end{figure}
		\newpage
		 As for partial susceptibilities, we should note that $\chi_{\mu,\mu} \geq 0,$ however $\chi_{\mu,\nu}$ with $\mu \neq \nu$  can be negative. Mark that in the low-temperature region $\chi_{0,1}$ suffers jumps at frustration points (see Figure \hyperref[hee]{11}) 	
				\begin{figure}[H]\label{hee}
			    \begin{minipage}[h]{\linewidth}
					\centering{\includegraphics[width=\linewidth]{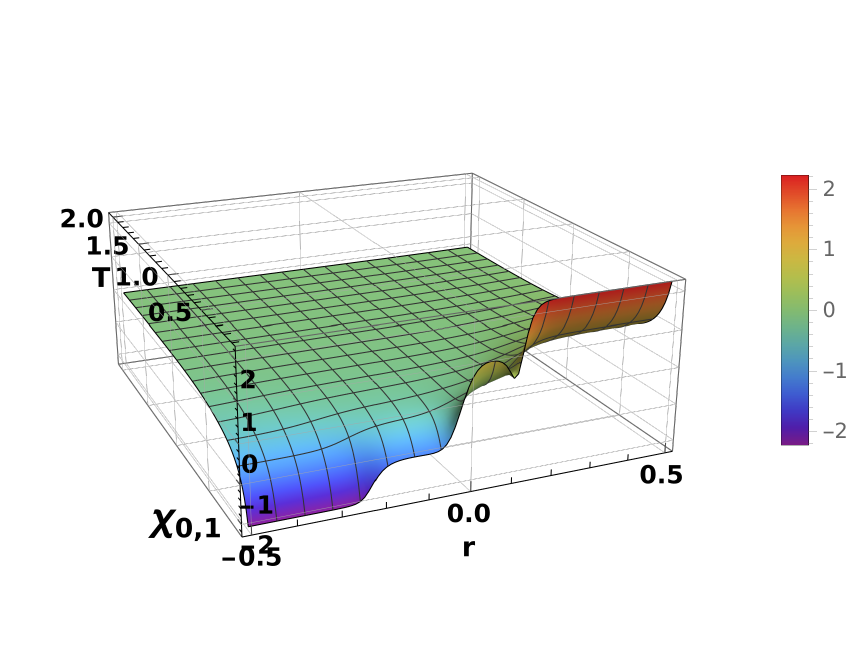} \\  }
				\end{minipage}
				\begin{minipage}[h]{0.495\linewidth}
					\centering{\includegraphics[width=\linewidth]{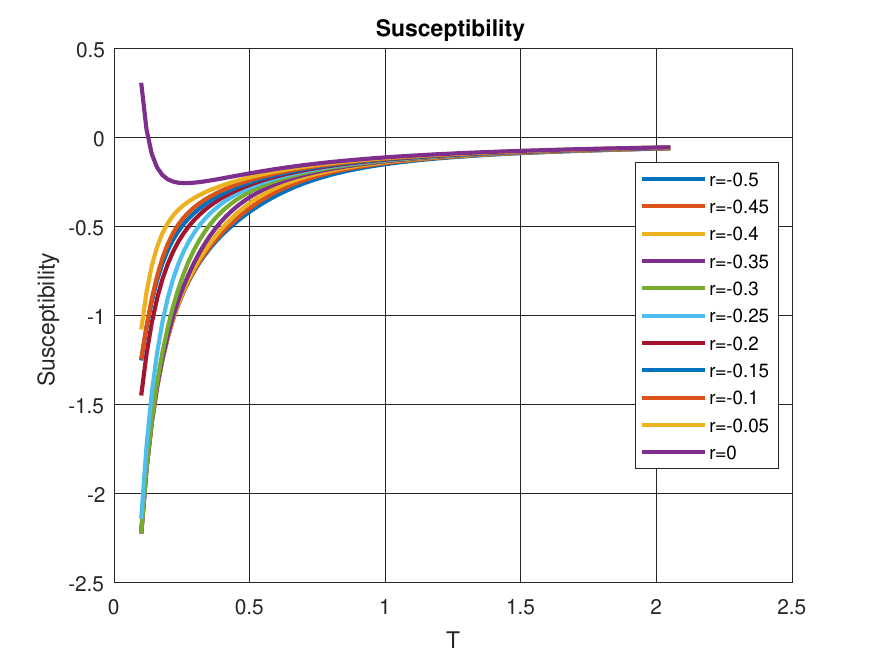} \\ }
				\end{minipage}
				\begin{minipage}[h]{0.495\linewidth}
					\centering{\includegraphics[width=\linewidth]{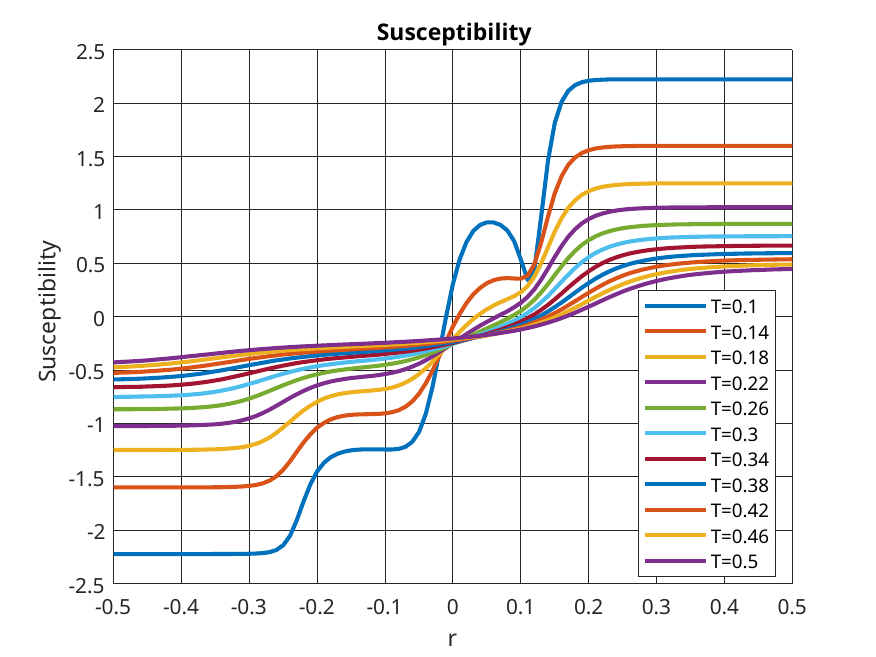} \\ }
				\end{minipage}
					\caption{ Plots of $\chi_{0,1}$, $r \in [-0.5;0.5],T \in [0.1;2.1]$} 			
				\end{figure}
				\newpage
	
	\section{Proof of theorems} \label{5}
	It is known that commuting matrices conserve eigensubspaces of each other \cite{Horn_Johnson}: $$AB=BA \Rightarrow    {AV_\lambda}   \subset  {V_\lambda},$$
where $V_\lambda-$ eigensubspace of matrix $B$. Indeed, if $AB=BA$ and $x$ is an eigenvector of matrix $B$, corresponding to eigenvalue $\lambda$, then vector $Ax$ is also an eigenvector of $B$, corresponding to eigenvalue $\lambda$. In other words, eigensubspaces of one operator must be invariant subspaces of another one.\\
\indent Let us formulate famous result from linear algebra: if linear space $\mathbb{V}$ can be decomposed into a direct sum of invariant subspaces of $\Theta$: $\mathbb{V}=\mathbb{W} \bigoplus \mathbb{Y} \bigoplus ...$, then spectrum of  $\Theta$ is equal to the union of spectra of  $\Theta$'s restrictions to its invariant subspaces: $\sigma (\Theta)=\sigma (\Theta \vert_{\mathbb{W}}) \bigcup \sigma (\Theta \vert_{\mathbb{Y}}) \bigcup ...$. Since generalized eigenvectors of $\Theta$ belongs to its invariant subspace, this means that search for generalized eigenvectors vectors of $\Theta$ can be performed not on the whole space, but on invariant subspaces, which are easy to find, because they are eigensubspaces of permutation matrix $\varkappa$ ( see (\ref{DD})). Finally, we obtain the form of the generalized eigenvectors:
\begin{equation}\label{D}
\begin {psmallmatrix} 
{v_1}& \\
{v_2}& \\
{v_3}& \\
{\varepsilon v_3}& \\
{\varepsilon v_1}& \\       
{\varepsilon v_2}& \\
{{ \varepsilon}^2 v_2}& \\
{{ \varepsilon}^2 v_3}&\\
{{\varepsilon}^2 v_1}&\\
\end{psmallmatrix}
\end{equation}
where $\varepsilon$ is a cubic root of 1.
\subsection{The form of the eigenvectors}
Permutation matrices permute the components of a vector. This fact allows us to find eigenvectors of these matrices \cite{Magret}. A permutation can be represented as a product of non-intersecting cycles. Let a permutation consist of only one cycle, for instance, $(123)$.
Then vectors of the form ${\left(1, \varepsilon, \varepsilon^2\right)}^T$, where $\varepsilon=\sqrt[3]{1}$,  are eigenvectors corresponding to the eigenvalue $\varepsilon$. When permutation is a product of several cycles one can easily find eigenvectors in a similar manner. 
For example, matrix  $\varkappa$ (\ref{kap}) corresponds to permutation $\left(1 5 9\right)\left(2 6 7\right)\left(3 4 8\right)$. Eigenvectors of matrix $\varkappa$, corresponding to eigenvalue $\lambda$, have the following form:
\begin{equation}\label{DD}
  \left.\begin{aligned}
  {(1,0,0,0, \varepsilon,0,0,0, {\varepsilon}^2)}^T\\
  {(0,1,0,0,0, \varepsilon,{ \varepsilon}^2,0,0)}^T\\
  {(0,0,1, \varepsilon,0,0,0,{ \varepsilon}^2,0)}^T
\end{aligned}\right\} \lambda= \varepsilon
\end{equation}

\section{Conclusion}
\indent In this paper, the new type of Potts model is introduced, which generalizes clock and standard models. In the clock interpretation, this model has 24 forces in a unit cell: 10 double, 8 triple and 6 quadruple ones. Suggested model covers all double-chain three-state Potts models with multi-site interactions which are invariant under cyclic shift of all spin values. An exact analytical solution of this model is obtained, formulae for its physical characteristics are given. Using the commutation of the transfer-matrix $\Theta$ with permutation matrix, the search of $9 \times 9$ matrix's spectrum is reduced to the search of spectruma of three matrices $3\times 3$ with some specific structure. Also  some generalized characteristics such as partial magnetization and susceptibility are introduced. An example of the model is analysed in detail: showed its ground states' structure and behaviour at low temperature.
\section{Appendix} \label{appendix}

\subsection{The Cardano formula}
To solve cubic equation $x^3+a x^2+b x+c=0$ at first one need (with the help of substitution $x+\frac{a}{3}=y$) to eliminate monomial $a x^2$ and get $y^3=-py-q$. Then if in $(d+e)^3=3de(d+e)+d^3+e^3,$ put $d+e=y$, we obtain a system:
\begin{equation*}
 \begin{cases}
  -q=d^3+e^3\\
   -p=3de
 \end{cases} \Longleftrightarrow
 \begin{cases}
 -q=d^3+e^3\\
 -\frac{p^3}{27}=d^3e^3
 \end{cases}
\end{equation*}
The last system can be easily solved. Let the pair of numbers $(d_0,e_0)$ be its solution. Then $(d_0{\varepsilon}^2, e_0\varepsilon), \ ( d_0\varepsilon,e_0{\varepsilon}^2)$ are also its solutions ( $\varepsilon$ is some complex cubic root of 1).
Eventually, we get the Cardano formula $$ y = \sqrt[3]{-\frac{q}{2}+\sqrt{\frac{p^3}{27}+\frac{q^2}{4}}}+\sqrt[3]{-\frac{q}{2}-\sqrt{\frac{p^3}{27}+\frac{q^2}{4}}}$$ \\

\end{document}